\documentclass[amsmath,amssymb,twocolumn,aps,pra,floatfix,superscriptaddress]{revtex4-2}

\usepackage{graphicx}
\usepackage{dcolumn}
\usepackage{bm}
\usepackage{hyperref}
\usepackage{float}
\usepackage[normalem]{ulem}
\usepackage{soul}
\usepackage{multirow}

\usepackage{color}
\usepackage[dvipsnames]{xcolor}

\begin{document}

\title{Benchmarking the role of particle statistics in Quantum Reservoir Computing
}

\author{Guillem Llodr\`{a}}
 \email{guillemllodra@ifisc.uib-csic.es}
\affiliation{%
 Institute for Cross-Disciplinary Physics and Complex Systems (IFISC) UIB-CSIC, Campus Universitat Illes Balears, Palma de Mallorca, Spain.
}
\author{Christos Charalambous}
\affiliation{%
 Institute for Cross-Disciplinary Physics and Complex Systems (IFISC) UIB-CSIC, Campus Universitat Illes Balears, Palma de Mallorca, Spain.
}
\author{ Gian Luca Giorgi}
\affiliation{%
 Institute for Cross-Disciplinary Physics and Complex Systems (IFISC) UIB-CSIC, Campus Universitat Illes Balears, Palma de Mallorca, Spain.
}
\author{Roberta Zambrini}
 \email{roberta@ifisc.uib-csic.es}
\affiliation{%
 Institute for Cross-Disciplinary Physics and Complex Systems (IFISC) UIB-CSIC, Campus Universitat Illes Balears, Palma de Mallorca, Spain.
}


\date{\today}

\begin{abstract}
Quantum reservoir computing is a neuro-inspired machine learning approach harnessing the rich dynamics of quantum systems to solve temporal tasks.  It has gathered attention for its suitability for NISQ devices, for easy and fast trainability, and for potential quantum advantage. Although several types of systems have been proposed as quantum reservoirs,  differences arising from particle statistics have not been established yet. In this work, we assess and compare the ability of bosons, fermions, and qubits to store information from past inputs by measuring linear and nonlinear memory capacity. While, in general, the performance of the system improves with the Hilbert space size, we show that also the information spreading capability is a key factor.
For the simplest reservoir Hamiltonian choice, and for each boson limited to at most one excitation, fermions provide the best reservoir due to their intrinsic nonlocal properties. On the other hand, a tailored input injection strategy allows the exploitation of the abundance of degrees of freedom of the Hilbert space for bosonic quantum reservoir computing and enhances the computational power compared to both qubits and fermions.
\end{abstract}
\maketitle
\normalsize

\section{Introduction}


Physical reservoir computing (RC) is a computational framework that harnesses a signal-driven dynamical system  to efficiently perform various machine-learning temporal tasks. 
Derived from the concepts of Echo State Networks (ESN) \cite{jaeger2001echo} and Liquid State Machines (LSM) \cite{maass2002real}, 
RC has the ability to map the input introduced into a higher-dimensional space. After this mapping,
only the readout layer needs to be trained to the expected output, while the reservoir is kept unchanged. This 
implies easy training and fast learning compared to traditional recurrent neural networks. Reservoir computers have proven successful in tasks such as spoken digit recognition \cite{verstraeten2005isolated}, continuous speech recognition \cite{triefenbach2014large}, and chaotic time series forecasting \cite{pathak2018model}. 
 Given the training simplicity, which does not require reservoir fine-tuning, any system able to develop a complex enough dynamics together with a fading memory with respect to inputs far in the past can be potentially used for RC purposes \cite{Konkoli2018ReservoirCW,tanaka2019recent,book_nakajima-Fischer}. Indeed, experimental implementations include photonic \cite{van2017advances}, electronic \cite{appeltant2011information} and spintronic \cite{torrejon2017neuromorphic}  set-ups. 


Still, an open question is which are the more relevant reservoir physical features conditioning the performance of an arbitrary RC scheme. 
In general, one expects that the size of the reservoir
state space, as well as the nonlinearity of the RC scheme, should play
a role. While in general the larger the space the better the resolution of differences in inputs, the relation between nonlinearity and memory, and how this affects
the RC performance is still not very well understood 
and is a major research direction in RC 
\cite{2012Lukovsevivcius,2015Goudarzi,book_nakajima-Fischer}.  
These fundamental questions become even more relevant once we move to the very recent field of quantum reservoir computing (QRC) \cite{Mujal2021}. Indeed,  compared to classical RC schemes, a major advantage is exactly identified in the richness of the Hilbert space where quantum particles reside.

QRC models have been proposed for both classical and quantum tasks  \cite{Mujal2021} and across a variety of different physical platforms.
Spin-based QRC systems, beyond the pioneering proposal of Ref. \cite{2017Fujii}, were also studied in Refs. \cite{PhysRevApplied.11.034021,Kutvonen2020,Chen2019,PhysRevApplied.14.024065,martinez2020information,Tran2021,2021Martinez,yelin,mujal2022time} under different lenses, while bosonic networks were proposed in \cite{nokkala2020gaussian} in the context of continuous-variable Gaussian states and  in Refs. \cite{PhysRevLett.123.260404,2021Govia,tureci,Kalfus2022}, outside the Gaussian limit, in the context of quantum extreme learning machines \cite{Mujal2021}. Finally, fermionic setups have been proposed for example in  classification tasks  like entanglement detection \cite{2019Ghosh} or for quantum circuit realization \cite{Ghosh_2021}. 

Despite the variety of such studies, often motivated by the experimental feasibility of the different platforms, what is missing is a comparative assessment of the (possible) different computational and learning capabilities inherited by quantum statistics.  For instance, it is known that in the context of quantum walk, fermionic and bosonic particles exhibit very different dynamical signatures and offer different degrees of robustness with respect to information localization \cite{sansoni}. The importance of quantum statistics 
has been also addressed in the context of non-equilibrium thermodynamics of quantum many-body systems\cite{Gong,Delgado,Poletti} or to assess the degree of coherence in the stationary state of open quantum systems \cite{Szameit}.
In this work, assuming as a model for QRC a quantum network of randomly coupled particles, we consider three distinct scenarios for quantum statistics, i.e. we study substrates made of three different types of particles, namely bosons, fermions, and qubits (or spin one-half particles). In order to assess the  computational power of each family, we will measure the linear and nonlinear memory capacities and separately analyze the effect of the system size, whose effective value we will find to be determined by the input injection strategy, and the effect of information spreading. 

We start in Sec. \ref{sec:methods} by introducing the model, the input injection strategy, the learning paradigm, and the performance quantifiers. In Sec. \ref{sec:results} we will present the main results of our work: linear and nonlinear memory capacity of the three kinds of physical particles; how the input-injection strategy can help improve the bosonic QRC performance;  and how nonlocal (anti-)commutation rules affect the information spreading in a quantum network.
Finally, the conclusions are presented in Sec.  \ref{sec:conclusions}.

\section{Methods}\label{sec:methods}

As discussed above, here we focus on the question of the role of the particle type on
the performance of quantum reservoirs. Namely, we will undertake
a comparative study on the performance of QRC whose substrates are
composed of bosons, fermions, or qubits. For the sake of simplicity, maintaining the main focus on the role of statistics,
we assume one of the simplest possible quadratic Hamiltonians:
\begin{equation}
H=\sum_{i,j=1}^{N}J_{ij}a_{i}^{\dagger}a_{j},
\label{eq:quadratic_hamiltonian}
\end{equation}
 where $J_{ij}$ are random number extracted from a uniform distribution in the interval $[0,1]$, $J_{ij}=J_{j,i}^*$, and $N$ is the number of particles. Here
we assume an all-to-all coupling among the particles. 
The operators $a_{i}$
($a_{i}^{\dagger}$), which correspond to the lowering (raising)
operators of the energy of the system can take any of the following three forms,
describing either fermions ($a_i\equiv f_i$), bosons ($a_i\equiv b_i$), or qubits, which are equivalent to $1/2$ spins ($a_i\equiv \sigma_i^-$).
The corresponding raising  operators are the relative conjugate ones.
We are going to consider a reservoir network of distinguishable particles.
These operators satisfy distinct commutation
relations, namely we have a not vanishing commutator for bosons 
$[b_{i},b_{j}^\dag]=\delta_{ij}$ 
while fermions satisfy the anticommutation relation $\{ f_{i},f_{j}^\dag\}=\delta_{ij}$, 
and  qubits,  according to the spin-$1/2$ algebra, obey $[ \sigma_{i}^a,\sigma_{j}^b]=2\delta_{ij}\epsilon_{abc}\sigma_{k}^c$, where $a,b,c$ are the spin-component indices  
and $\epsilon_{abc}$ is the Levi-Civita symbol. In the following, we will refer to either qubits or spins equivalently.

In order to inject information into the reservoir, we follow the  encoding  proposed in Ref. \cite{2017Fujii}. At   regular intervals of time $\Delta t$, 
inputs  belonging to a uniform random time series $\{ u_{k}\} _{k=1}^{L}$ ($0\leq u_k \leq1$) 
are  introduced to the system
sequentially by updating the state of one particle (let us say particle $1$) according to
\begin{equation}
|\psi_{k}^{(e)}\rangle =\sqrt{u_k}\left|0\right\rangle +\sqrt{1-u_k}\left|e\right\rangle, \label{eq:input_state}
\end{equation}
where the state label $e$ will be assumed to be $e=1$ for the case of  fermions and qubits (where we can map $|0\rangle \to  |{\downarrow}\rangle$ and $|1\rangle \to  |{\uparrow}\rangle$), while $e\in\mathbb{Z}^+ $
for bosons, representing the energy level of the input particle that
we populate at each new data injection. 
 We anticipate that the bosonic infinitely large Hilbert space poses a challenge to numerical simulations, that, as usual, are faced with truncation at an energy level affecting the least the description.  
The introduction of  inputs (Eq. (\ref{eq:input_state})) is assumed to be
instantaneous and for the rest of the time, the reservoir evolves under the quantum dynamics of a closed system, namely the unitary evolution associated with the Hamiltonian in Eq. (\ref{eq:quadratic_hamiltonian}).  As a result,  the completely positive trace preserving (CPTP) map describing the reservoir dynamics is 
\begin{equation}
\rho\left(k\Delta t\right)=e^{-iH\Delta t}\left[\rho_{1,k}^{(e)}\otimes {\rm Tr}_{1}\left\{ \rho\left(\left(k-1\right)\Delta t\right)\right\} \right]e^{iH\Delta t},
\label{eq:reservoir_evolution}
\end{equation}
with ${\rm Tr}_{1}\left\{ \cdot\right\} $ denoting the partial trace performed
over the first particle, where the input has been introduced, and with $\rho_{1,k}^{(e)}=|\psi_{k}^{(e)}\rangle \langle \psi_{k}^{(e)}|$. 
Finally, as in previous works, information is extracted from the quantum system by measuring a number of observables, which define the output layer. 
The set
of observed nodes at time $k\Delta t$ is defined as $\left\{ x_{j}\right\} _{j=1}^{M}$,
where 
\begin{equation}
x_{j}\left(k\Delta t\right)={\rm Tr}\left[O_{j}\rho\left(k\Delta t\right)\right],
\end{equation}
and where $O_{j}$ is the $j^{th}$ observable chosen from a list of $M$
elements. 
Notice that for  the particular Hamiltonian we are considering,  varying $\Delta t$ is equivalent to changing the coupling strength  range $J$  among the particles. 
Also notice that the number of computational nodes entering the learning process can be enlarged, even beyond the number of independent observables for each reservoir, using the temporal multiplexing technique, through which the signals
are sampled  not only at the time $k\Delta t$, but also
at each of the subdivided $V$ time intervals during the unitary evolution,
resulting in the construction of $V$ virtual nodes \cite{2017Fujii,martinez2020information}. As discussed in more detail in Ref. \cite{mujal2021analytical}, this  scheme produces a nonlinear input-output mapping where the encoding plays a key role.

Another point to be made is that here we consider an ensemble quantum system made  
of a huge number of copies of the reservoir. That is, we neglect the effect of measurement back-action and refer the reader to Ref. \cite{mujal2022time} for a detailed discussion about this aspect of QRC.

As for the learning stage,  
let $\left\{ x_{j}\left(k\Delta t\right)\right\} $
where $1\le j\le M$ and $1\le k\le L$, be the set of the average values of the $M$ observables at $L$ time steps mentioned above. Furthermore, let $\left\{ \widehat{y}\left(k\Delta t\right)\right\} _{k=1}^{L}$
be the target sequence  that we want to emulate and let $y_{k}=F\left(\left\{ u_{k}\right\}\right)$,
where $F$ encodes the function transforming the input into the outcome  of the measuring process.  
Since the quantum reservoir dynamics we are considering already introduces nonlinearity in the process \cite{mujal2021analytical}, the outcome read-out function can be simply assumed to be a linear function of the measurements, as usual:
\begin{equation}
y_{k}=\textbf{w}\textbf{x}\left(k\Delta t\right) +\textbf{b},
\end{equation}
where $\textbf{w}=\left\{ w_{j}\right\} _{j=1}^{M}$ is a vector with the readout
weights of the linear output function and $\textbf{b}$ a constant bias term. Hence in the reservoir computing
approach, learning of a nonlinear function $y_{k}$ which emulates
the target sequence $\widehat{y}$, amounts simply in training the
linear readout weights of the reservoir states. This training
is done by minimizing the deviation from the target, which is usually chosen to be quantified
through the mean-square error, such that
\begin{equation}
arg\min_{\left\{ w_{j}\right\} _{j=1}^{M}}\sum_{k}\left(\widehat{y}_k-y_{k}\right)^{2}.
\end{equation}
There are a number of ways one can solve this optimization problem.
The easier and most common one is linear regression, where one needs to first
re-express the problem in a matrix form, where the measured function reads as
\[
\mathbf{y}=\mathbf{W_{out}}\mathbf{X}.
\]
The
weights that minimize the mean-square error are then determined by the
Moore-Penrose pseudo inverse $W_{out}^T=y^{\dagger}\mathbf{X}$.  Essentially, the training phase for a reservoir computer
requires only a single matrix inversion, hence, it offers considerable
computational savings over traditional neural networks \cite{book_nakajima-Fischer}. 
A diagrammatic description of the whole QRC protocol described here can be found in Table.~\ref{table:diagrammatic_scheme}. This will be considered for reservoirs of different particles.
\begin{table}[H]
    \centering
    \begin{tabular}{cc} 
    \hline \hline
    \multirow{2}{*}[-0.2cm]{\textbf{Input}} & $|\psi_{k}^{(e)}\rangle =\sqrt{u}_{k}|0\rangle +\sqrt{1-u_{k}}|e\rangle$
    \rule[-0.45cm]{0pt}{0.9cm} \\ 
    & $\rho_{1,k}^{(e)}\otimes {\rm Tr}_{1}\left\{ \rho\left(\left(k-1\right)\Delta t\right)\right\}$  
    \rule[-0.3cm]{0pt}{0.6cm} \\
    \hline
    \multirow{2}{*}[-0.2cm]{\textbf{Reservoir}} & $H=\sum_{i,j=1}^{N}J_{ij}a_{i}^{\dagger}a_{j}$
    \rule[-0.45cm]{0pt}{0.9cm} \\
    & Eq. (\ref{eq:reservoir_evolution})  dynamics
    \rule[-0.3cm]{0pt}{0.6cm}\\
    \hline
    \multirow{2}{*}[-0.2cm]{\textbf{Output}} & $x_{j}\left(k\Delta t\right)={\rm Tr}\left[O_{j}\rho\left(k\Delta t\right)\right]$
    \rule[-0.45cm]{0pt}{0.9cm} \\ 
    & $y_{k}=\textbf{w}\textbf{x}\left(k\Delta t\right)$
    \rule[-0.3cm]{0pt}{0.6cm} \\
    \hline \hline
    \end{tabular}    
\caption{\label{table:diagrammatic_scheme} Schematic representation of how QRC processes information. (a) At each iteration the input value ($u_k$) is encoded into a quantum state,  $|\psi_{k}^{(e)}\rangle$ and injected into one of the particles of a many-body system (reservoir). The dynamics is ruled by a quadratic Hamiltonian and the map Eq. \eqref{eq:reservoir_evolution}. The output layer ($x_j$) is accessed through the measurement of a set of observables ($O_j$), whose linear combination is optimized to produce the target associated with each task.} 
\end{table}

Finally, let us introduce a quantifier for the QRC capabilities. By convention, the system performance, that is, the computational capacity for different tasks is quantified  using the coefficient of determination (or Pearson correlation)
between the output $y_{k}$ and the desired output $\widehat{y}_{k}$

\begin{equation}
C\text{=}\frac{{\rm Cov}^{2}\left(\widehat{y}_{k},y_{k}\right)}{\sigma^{2}(y_{k})\sigma^{2}(\widehat{y}_{k})},\label{eq:MC DELAY}
\end{equation}
where $\sigma(\widehat{y}_{k})$ and $\sigma(y_{k})$  are the standard deviation of the
target output and reservoir output over time, respectively and ${\rm Cov}$ indicates the covariance. 

\section{Results}\label{sec:results}
The main goal of this paper is to compare the performance of particles
of different statistics. Before, 
it is crucial to establish certain operating features
of our setups. To begin with, we need to check that a reservoir composed of bosons with a truncated Hilbert
space is a sufficiently faithful description of the numerically inaccessible  system
of the bosons with infinite-dimensional Hilbert spaces. To do this, we need to check the population of energy levels above the selected cutoff $n_{\rm co}$, to make sure that these levels do not play any important role in the dynamics and therefore in the performance of the QRC. This test also enables us to identify at which point of the energy spectrum
of the Hilbert space we should truncate. 
\begin{figure}[ht]
    \centering
    \includegraphics[width=0.48\textwidth]{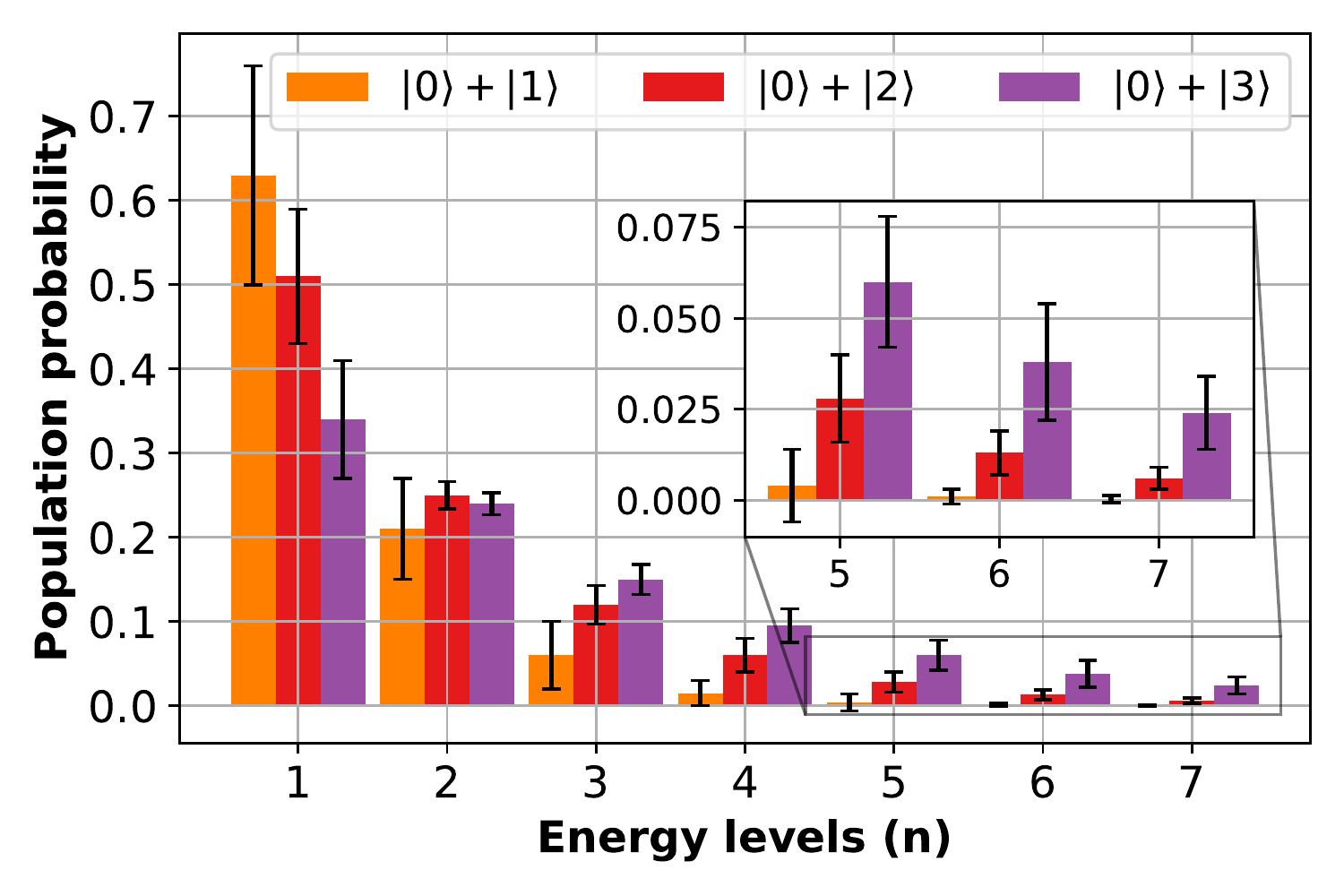}
    \caption{\label{fig:Cutoff} Probability to find bosons at different energy levels $n$. Given a reservoir with four bosons, the injected quantum state $|\psi_{k}^{(e)}\rangle$ can excite higher energy modes depending on the encoding $e=\left\{1, 2, 3\right\}$. Only the first two encodings allow us to accurately truncate the Hilbert space $n_{co}=5, 6$ for $e=1, 2$, respectively). For encoding $e=3$, the chance of finding bosons at $n=7$ is still above $1\%$, so this encoding would not be accurately described by this truncation. 
    The results are computed following Eqs. (\ref{eq:quadratic_hamiltonian}-\ref{eq:reservoir_evolution}) with $\Delta t=10$ and averaged over 100 realizations of the reservoir network ($J_{ij}$) and of input sequences $u_k$. Error bars represent the standard deviation.}
\end{figure}
We notice that the dynamics considered here contains dissipation, involving a partial trace at each injection step \eqref{eq:reservoir_evolution}, but is at the same time driven by the external input that could in principle introduce high excitations $e$ into the reservoir \eqref{eq:input_state}.
As we can see in Fig. \ref{fig:Cutoff}, for the input state $|\psi_{k}^{(1)}\rangle $ and $|\psi_{k}^{(2)}\rangle $ the energy levels can be safely cut off at $n_{\rm co}=5$ and $n_{\rm co}=6$ respectively, being these levels negligibly populated. Actually, the truncation at those energy levels is justifiable as well as numerically achievable for a 4 bosons reservoir, even when considering the computation of QRC for several tasks and reservoir realizations as in this work. 

\begin{figure}[ht]
    \includegraphics[width=0.48\textwidth]{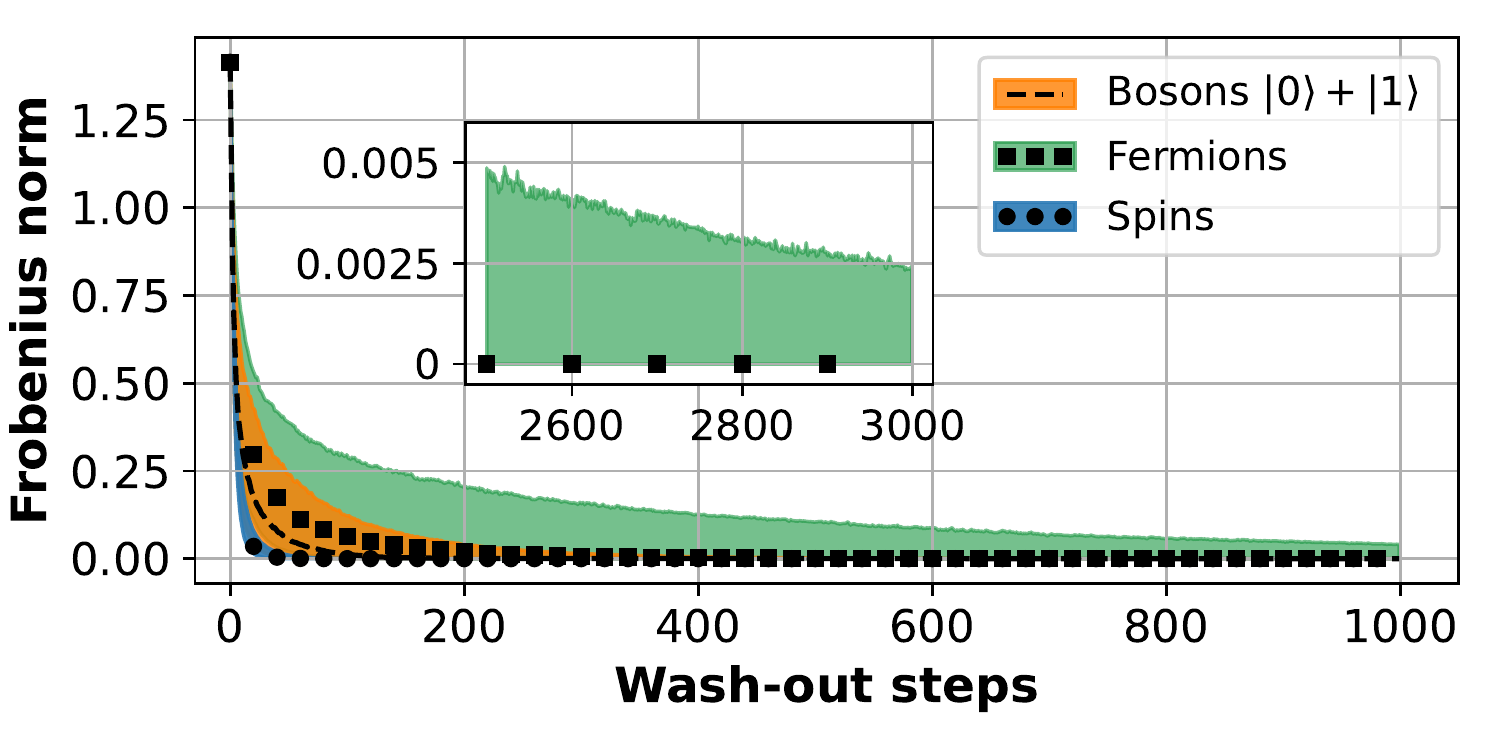}
     \caption{\label{fig:echo_property_with_median} Number of iterations to satisfy the convergence property for different particles. The Frobenius distance is computed for reservoirs with $N=4$ particles with initial states $\rho_{A}$ and $\rho_{B}$ as defined in the main text. The dotted lines represent the median over 1000 realizations of reservoirs and input sequences and the shaded area represents the first and third quantiles. All particles satisfy the convergence property for sufficiently long times. Qubits and bosons quickly converge to zero, while fermions remember the initial conditions longer.}
     \end{figure}

A second aspect we need to assess for the proper QRC operation is whether our reservoir computer is equipped with the convergence or echo state
property, which means that, after repeated input injection,  the reservoir forgets its initial state. This property is fundamental for good performance \cite{jaeger2001echo} and depends on the features of the QRC evolution map.
Convergence is proven by studying, after a sufficiently long time,
the distance between two  states of the reservoir evolving from two  distinct 
initial conditions. In particular, we plot in  Fig. \ref{fig:echo_property_with_median} the Frobenius distance $\Vert A\Vert =\sqrt{{\rm Tr} (AA^{\dagger})}$
between the states $\rho_{A}=|0\rangle \langle 0|\otimes|1\rangle \langle 1|\otimes|0\rangle \langle 0|\otimes|0\rangle \langle 0|$
and $\rho_{B}=|0\rangle \langle 0|\otimes|0\rangle \langle 0|\otimes|1\rangle \langle 1|\otimes|0\rangle \langle 0|$,
for bosons, fermions, and for spins. In all cases, the distance between states decays after a few hundred of steps and this  convergence time will be used in the following tasks of QRC (to set the  initial time for training). However, we see that fermions, contrary to bosons and qubits, take a much longer time to forget, and we verified that this was independent of the initial state choice. This is in agreement with the fact that correlations between distinct sites $\langle a_{i}^{\dagger}a_{j}\rangle $ for $i\neq j$ are in general larger for fermions compared to the other two cases, indicating the capacity of the fermionic system to store information more non-locally compared to bosons or qubits, and entailing a major resilience in forgetting the initial state. We will further discuss this feature in Subsec. \ref{subsec:compare}.  Finally, we observed a much larger error bar (filled area) among the realizations of reservoir random networks and input series for fermions (green area) compared to that of bosons (orange) and qubits (blue). 
In the following tasks, 
we will set the number of iterations to remove the initial transient (wash-out) 
to 1000 iterations for bosons/spins and to 3000 for fermions.

\subsection{Tasks}\label{subsection: Tasks}
We will now assess the performances of the three QRC setups by considering tasks requiring linear and non-linear memory capacity. The numerical code created for the Figure \ref{fig:linear_delay} task is publicly available
\cite{github}.

\subsubsection{Linear memory capacity}
The linear memory capacity (MC)
is a standard measure of memory in recurrent
neural networks. It is defined as the capacity $C$ defined in Eq.
(\ref{eq:MC DELAY}) when  the target function is
\begin{equation}
\widehat{y}_{k}=u_{k-\tau},
\end{equation}
where $\tau\in\mathbb{N}$ is the delay. Note that $\tau\geq0$ and
hence this measure can serve to describe how well the task of reconstructing
the delayed version of the input is performed. 
The plots presented here are the averages of 1000 runs. The default set of observables used as the output layer (to start with) is
${\rm Re}[\langle a_{i}^{\dagger}\left(k\right)a_{j}\left(k\right)\rangle] $, but we will see later that this can be extended to other ones.

\begin{figure}[ht]
    \centering
    \includegraphics[width=0.48\textwidth]{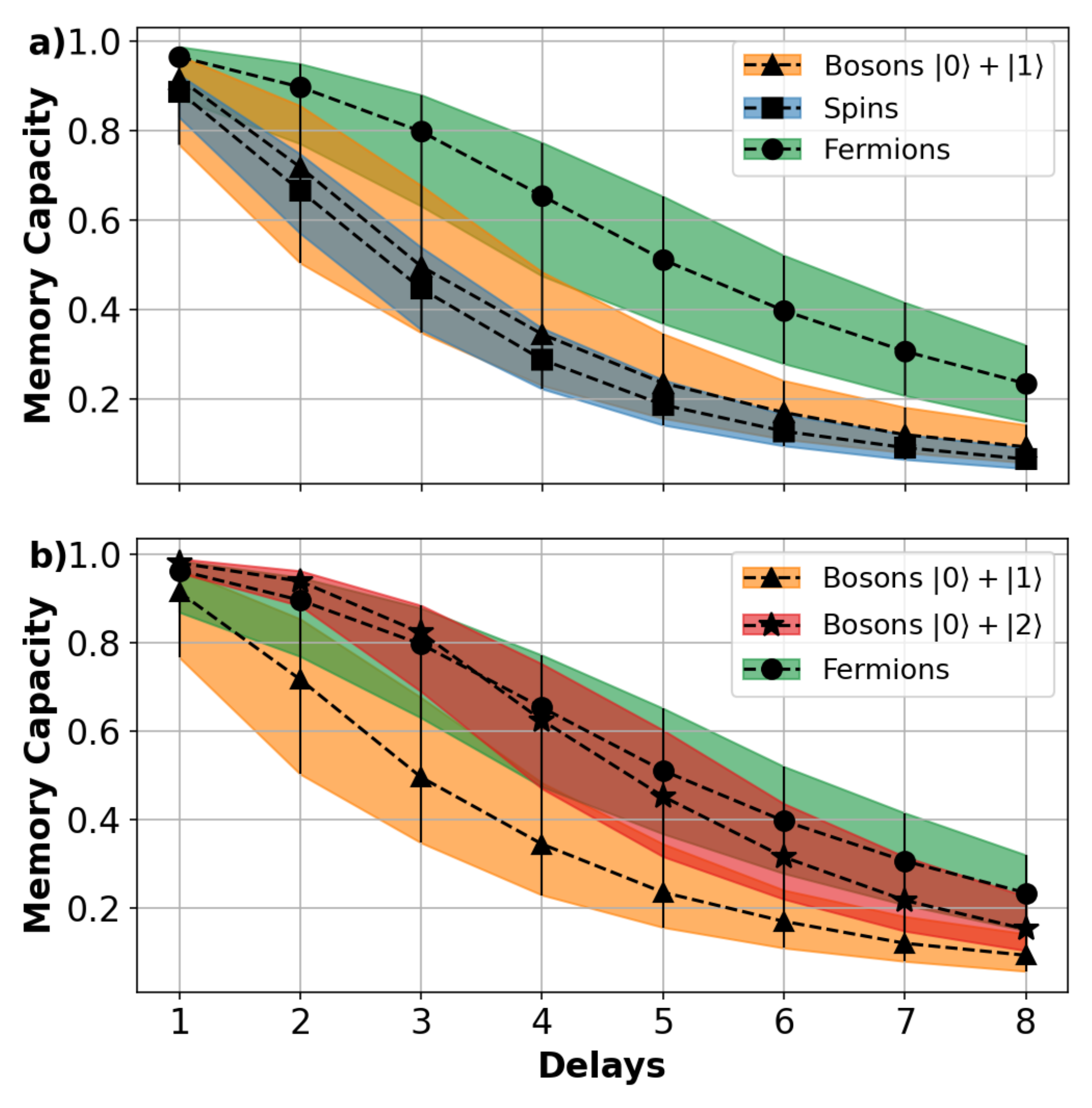}
    \caption{\label{fig:linear_delay} Linear MC as a function of the input delay for different (a) particles and (b) encodings. (a) Fermions better recall past inputs than spins and bosons. Surprisingly, a bosonic reservoir with encoding $e=1$ gives a similar performance to a spin reservoir. (b) However, if bosons are properly excited ($e=2$) there is a significant boost in their performance. As denoted by the error bars the memory capacity is extremely sensitive to the values of $J_{ij}$. For this reason, the number of realizations has increased from 100 to 1000, with respect to Fig. \ref{fig:Cutoff}. The dotted lines represent the median, and the error bars indicate the first and third quantiles. To avoid overfitting the learning algorithm is trained over $L=1200$ samples of $u_k$ and tested over $L=300$ new samples (a-b). The remaining parameters are $N=4$, $\Delta t=10$, $M=10$ (the output observables are $\langle a_{i}^{\dagger}a_{j}\rangle$), $n_{\rm co}=5$ ($n_{\rm co}=6)$ for bosons with $e=1$ ($e=2$) and a wash-out of 1000 (3000) iterations for bosons/spins (fermions).}
\end{figure}
In Fig. \ref{fig:linear_delay} we study the dependence of  the memory capacity (MC) on the delay. We see that as we increase the delay, as  expected, the memory capacity drops in all cases, indicating that the reservoir is better at remembering more recent inputs. In Fig. \ref{fig:linear_delay}a, 
we plot the cases of a fermionic reservoir, a qubit reservoir, and a bosonic one where the input state is introduced using the first excited state of the input boson ($e=1 $ in Eq. (\ref{eq:input_state})). On the other hand, in Fig. \ref{fig:linear_delay}b, we focus on the behavior of fermions with respect to bosons where the input state is introduced  either in the first or in the second excited state.  

By comparing these two figures, we see that a bosonic reservoir computer where only the first state is excited through the input injection performs approximately as well as a reservoir composed of qubits and worse than a fermionic reservoir. This result can be explained as follows: since only the first excited state of the input boson is considered, the energy pumped into the setup is so low that in practice only the lower two dimensions of any of the bosons are explored. Then, the abundance of levels of the Hilbert space does not play any constructive role, as only a few of the density matrix elements have nonzero entries. In this case, the poor number of excitations dresses the bosonic commutation rules in such a way that, when applied to the reservoir density matrix, these are effectively indistinguishable from the spin commutation rules (a double excitation operator will have matrix elements practically equal to zero over the reservoir state).  Hence, bosons and spins perform similarly. 
What is more, in this figure, is also observed how fermions perform better than spins or bosons with the first excited input state. We attribute this to the fact that fermions, due to the anticommutation relations, are capable of storing information more non-locally compared to spins
and bosons. On the other hand, in the bottom of Fig. \ref{fig:linear_delay},
we see that a bosonic reservoir computer with higher input excitations (even if only up to $e=2$ in Eq. \eqref{eq:input_state}) reaches the performance of a fermionic reservoir. We attribute this
improvement to the larger Hilbert space available to bosons. 
If the third excited state  were used to encode the input, bosons would presumably  perform
even better, but simulating this setup with the needed accuracy requires considering larger energy cutoffs, which is computationally very expensive. Also, one has to take into account that performances can quickly saturate as the Hilbert space size increases if the dynamics cannot make all the degrees of freedom linearly independent with respect to each other \cite{Kalfus2022}.

\begin{figure}[t]
    \centering
    \includegraphics[width=.48\textwidth]{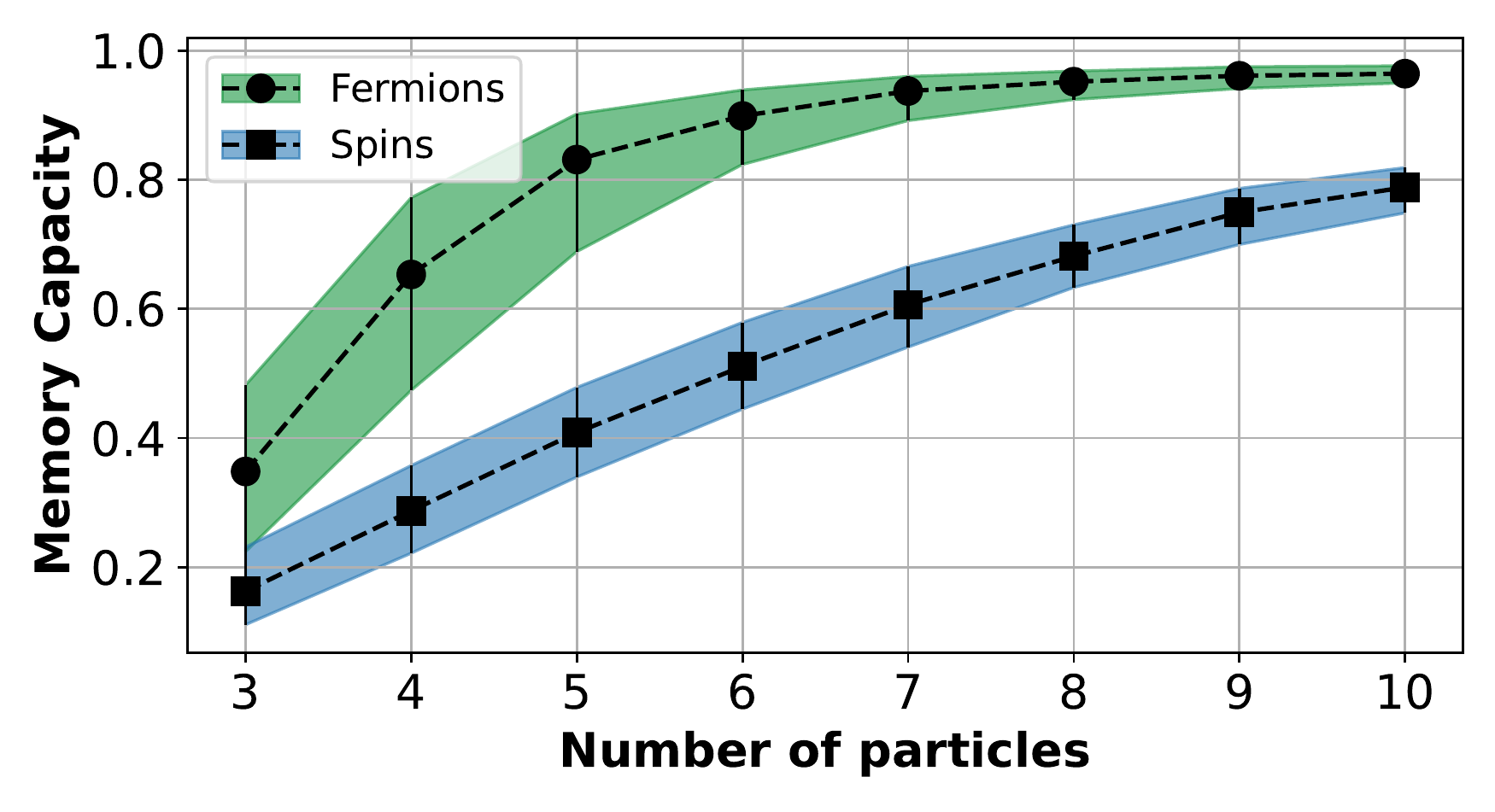}
    \caption{\label{fig:linear_parameters} Dependence of the linear MC on the number of particles for the case of fermions and spins. The delay is fixed at $\tau=4$. The results presented are obtained using the same parameters as in Fig \ref{fig:linear_delay}, including the number of realizations.}
\end{figure}

In Fig. \ref{fig:linear_parameters} we show the dependence of MC
on the system size for fermions and qubits for a fixed value of the delay ($\tau=4$). We see that increasing the number of particles improves the performance of the reservoir as expected, but the rate of increase is significantly larger for fermions. This result is a further confirmation of the favorable dynamics of fermions that we relate to the higher spreading information capacity of fermion networks, with respect to the case of qubits.

\begin{figure}[h]
    \centering
    \includegraphics[width=0.48\textwidth]{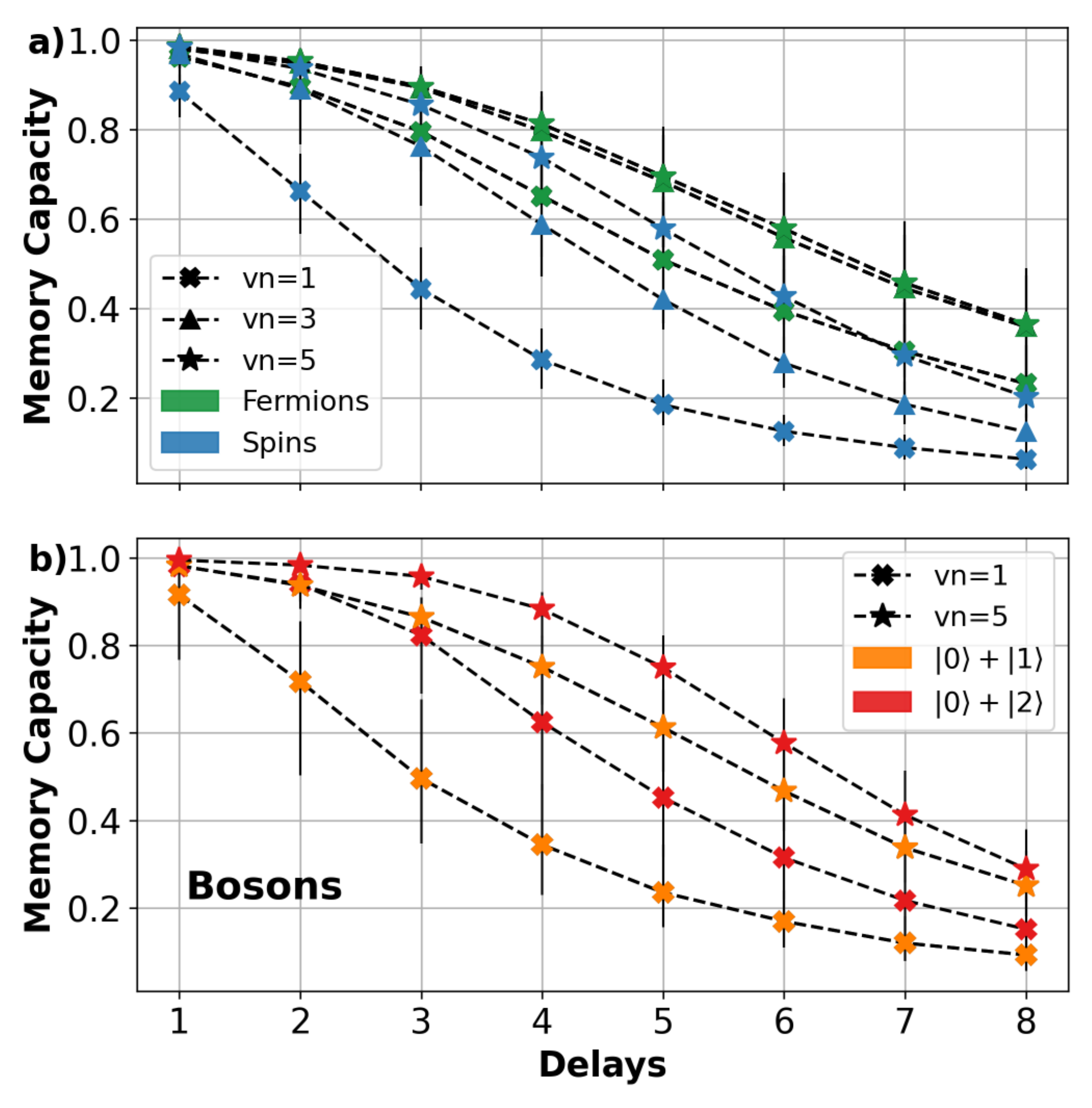}
    \caption{\label{fig:linear_virtual} Linear MC with different number of virtual nodes ($vn$) for (a) fermions-spins and (b) bosons with $e=1, 2$ (see Eq. \eqref{eq:input_state}). 
    In general, the temporal multiplexing technique improves the performance in all cases. (a) Fermions saturate their performance with 3 virtual nodes. (b) When $e=2$, bosons still perform better than in the case $e=1$. Additionally, when $e=2$ bosons can outperform fermions for short delay ($\tau \leq 4$). As the number of virtual nodes increases, bosons and spins tend to overfit. To solve this problem, for $vn \geq 3$ we have simply increased the amount of data up to $L=2000$ ($L=500$) for the train (test) data. Again the dotted lines represent the median and the error bars the first and third quantiles, all remaining parameters can be found in Fig. \ref{fig:linear_delay}.}
\end{figure}

A resource for QRC is not only the reservoir size but also a large output layer, including virtual nodes, as commented before. In Fig. \ref{fig:linear_virtual} we study the dependence of MC on
the number of virtual nodes for fermions, bosons, and qubits. We see
that in all cases, introducing virtual nodes improves the performance,
but for fermions, this leads to saturation faster than for spins. 
We also see that, in the case of bosons with the input accessing excited levels ($e=2$), virtual nodes further enhance the reservoir performance. This points to the fact that dealing with a more complex dynamics, exciting higher levels $n$, provides the larger space needed to improve QRC performance (also beyond the mere fact of having more output degrees of freedom). 

\begin{figure}[b]
\includegraphics[width=0.48\textwidth]{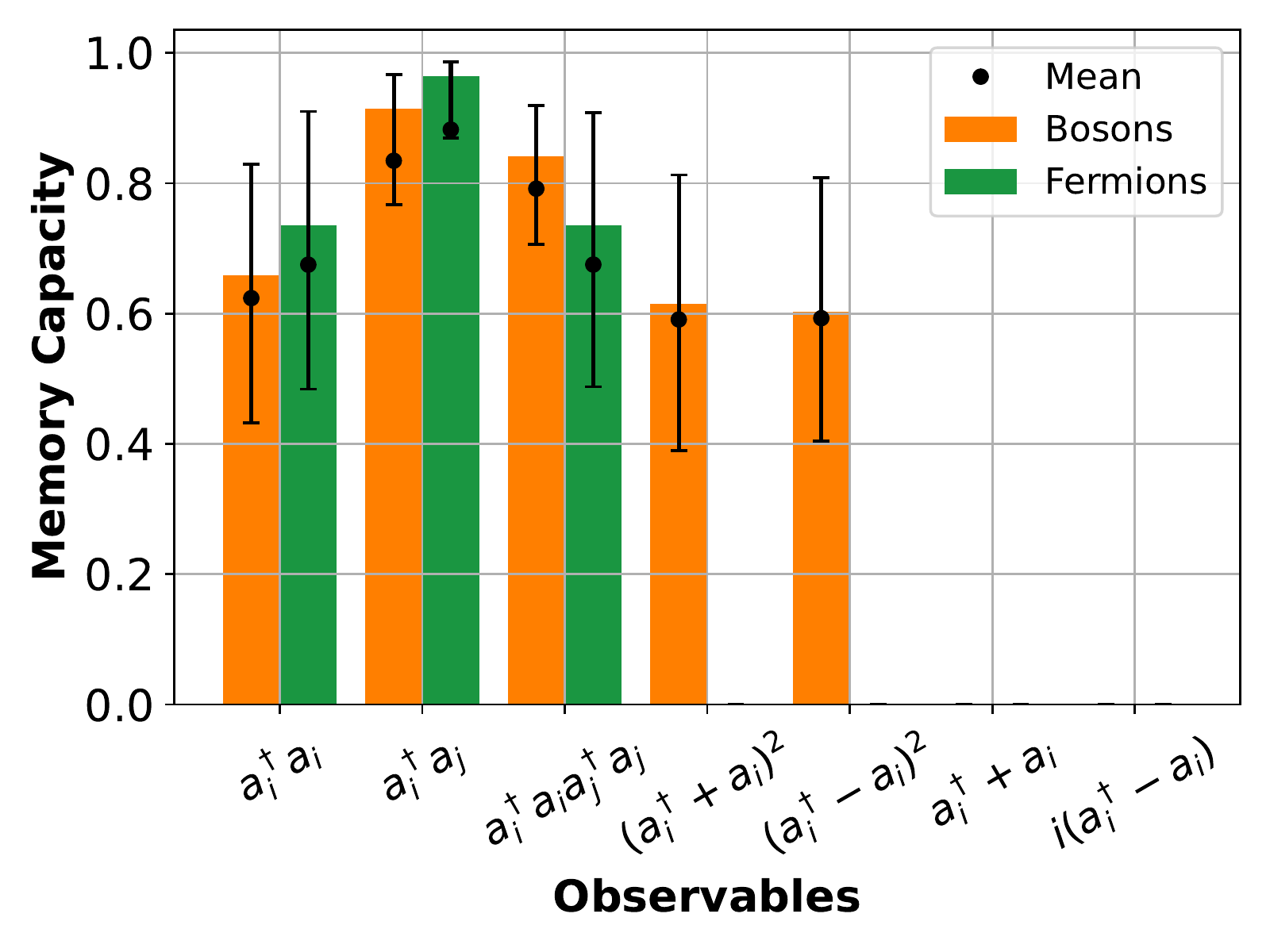}
\caption{\label{fig:linear_observables} 
Linear MC for different observables. The delay is fixed at $\tau=1$ and $vn=1$. Usually, the median (bar height) and the average (dot) will not match because of the presence of outliers, while the error bars indicate the first and third quantiles.
If we consider either occupation numbers or the (symmetrized) operators $\langle a_{i}^{\dagger}(k) a_{j}(k)\rangle $ as output layer entries,
we see that both types of particles have about the same capacity for
functioning as QRC, with the fermions performing slightly better.
On the contrary, when higher-order cross-correlations are considered as the observables,
i.e. terms of the form $\langle a_{i}^{\dagger}(k)a_{i}(k) a_{j}^{\dagger}(k) a_{j}(k)\rangle $,
bosons improve their capacity, performing better than fermions as
QRC. }
\end{figure}

To conclude the linear memory analysis, we plot in Fig. \ref{fig:linear_observables} the effect of considering
different observables by comparing bosons and fermions. 
The first
point to note here, is that both for the observables $\langle n_{i}\rangle=\langle a_{i}^{\dagger}a_{i}\rangle $,
i.e. the occupation numbers, and for the cross terms $\langle a_{i}^{\dagger}a_{j}\rangle $,
$i\neq j$ the capacity for both fermions and bosons is similar, with
fermions performing slightly better. 
We find that higher
order cross terms, $\langle a_{i}^{\dagger}a_{i}a_{j}^{\dagger}a_{j}\rangle $, do not reach the best performance, being the non-diagonal correlations  $\langle a_{i}^{\dagger}a_{j}\rangle $ the most useful in terms of linear memory capacity. This justifies our choice of the output layer observables in the rest of the manuscript. We also notice that for the observables $\langle a_{i}^{\dagger}+a_{i}\rangle $
and $\langle a_{i}^{\dagger}-a_{i}\rangle $ both fermions
and bosons perform quite badly. Finally, for bosons
the observables $\langle (a_{i}^{\dagger}\pm a_{i})^{2}\rangle =\langle (a_{i}^{\dagger})^{2}\rangle +\langle a_{i}^{2}\rangle \pm(\langle a_{i}^{\dagger}a_{i}\rangle +\langle a_{i}a_{i}^{\dagger}\rangle )=\langle (a_{i}^{\dagger})^{2}\rangle +\langle a_{i}^{2}\rangle \pm(2\langle a_{i}^{\dagger}a_{i}\rangle +1)$
give a MC of the order of the diagonal contributions $\langle a_{i}^{\dagger}a_{i}\rangle$, 
On the other hand, for fermions, these quantities lead to a vanishing MC. This can be easily understood as  $\langle (a_{i}^{\dagger}\pm a_{i})^{2}\rangle =\langle (a_{i}^{\dagger})^{2}\rangle +\langle a_{i}^{2}\rangle \pm(\langle a_{i}^{\dagger}a_{i}\rangle +\langle a_{i}a_{i}^{\dagger}\rangle )=\pm 1$. Hence, the dynamics is not able to capture the necessary features to recall past inputs. Similarly, the dynamics of observables $\langle a_{i}^{\dagger}+a_{i}\rangle$
and $\langle a_{i}^{\dagger}-a_{i}\rangle $ is not rich enough to distinguish different input injections. This can be directly visualized by looking at the real-time dynamics of the observables (see Appendix \ref{appendix:dynamics}.

\subsubsection{Non-linear tasks}\label{subsec:non-linear tasks}

Let us now study the capability of our system to remember a nonlinear function of an input injected in the past: in this case, the target is given as
\begin{equation}
\widehat{y}_{k}=u_{k-\tau}^{q}
\end{equation}
where $\tau\in\mathbb{N}$ is the delay and $q\in\mathbb{N},q>1$ is
the degree of the polynomial nonlinear function of the input. 
The results
are for averages over $1000$ realizations and the default observable used
is ${\rm Re }[\langle a_{i}^{\dagger}(k)a_{j}(k)\rangle ]$.

\begin{figure}[h]
    \centering
    \includegraphics[width=0.48\textwidth]{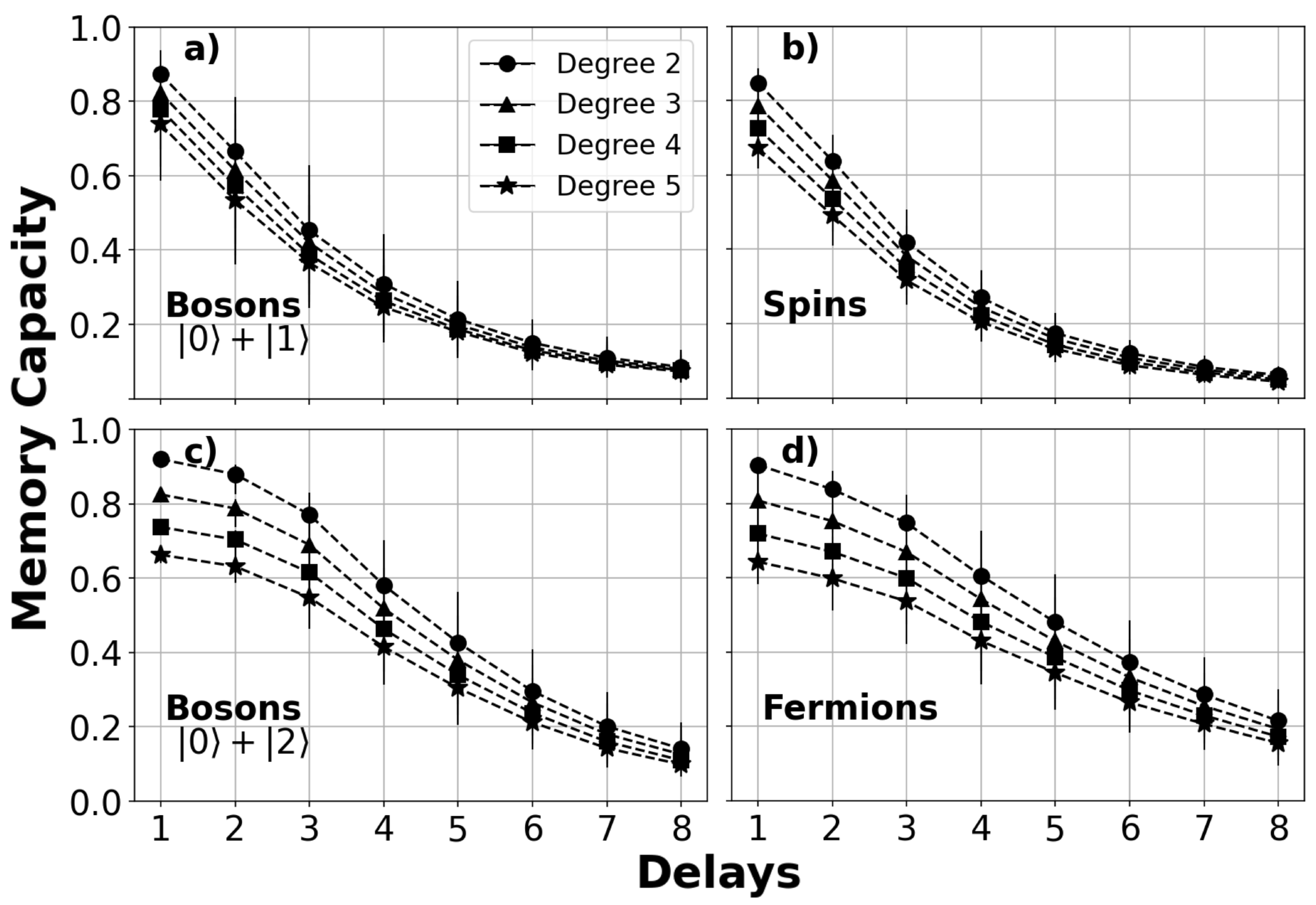}
    \caption{\label{fig:non_linear} Non-linear MC as a function of the delay with different degree of non-linearity ($q=2$ to $q=5$). (a-b) As in the linear case, bosons with $e=1$ behave similarly to spins, (c-d) while bosons with $e=2$ behave similarly 
 to fermions. Also in this case,  all parameters are identical to the ones used in  Fig. \ref{fig:linear_delay}.}
  \end{figure}

In Fig. \ref{fig:non_linear}, we plot the performance of the QRC as a function of the delay for four distinct setups: bosons with the input state being a combination of the ground state and either the first (Fig. \ref{fig:non_linear} top left) or second (Fig \ref{fig:non_linear} bottom left) excited state, fermions, and spins. For all these cases, we plot the performance for tasks of different non-linear degrees depending on the delay. We see that in all four cases, increasing the nonlinear degree reduces the performance of the QRC, that is highly non-linear memory is harder to achieve. We also see again that bosons with the first excited state in the input perform similarly to qubits, while bosons with $e=2$ excitations perform similarly to fermions. Therefore, all the results obtained for the linear memory capacity are qualitatively equivalent to the ones of the nonlinear case.

While here we have analyzed the ability of our system to process a given linear or nonlinear function of the input,  a  powerful  estimation of the performance is given by the information processing capacity (IPC), introduced in Ref. \cite{2012Dambre}, accounting for \textit{all} possible nonlinear functions of the input as targets
(taking a complete family, such as Legendre polynomials). In Appendix \ref{appendix:IPC}, we calculate the IPC profile for fermions, spins, and bosons and make a  comparative analysis with respect to  the MC, showing the similarities and differences between these two kinds of approaches.

\subsection{Comparing fermions and qubits}\label{subsec:compare}

One of our main results is that fermions are more suitable than qubits to store information over the past inputs Fig. \ref{fig:linear_delay}. To explain why fermions perform better than qubits, we consider here 
a toy model to illustrate how  the anticommutativity nature of fermions can lead to a better spread of the information through the whole reservoir.

\begin{figure}[H]
    \centering
    \includegraphics[width=0.48\textwidth]{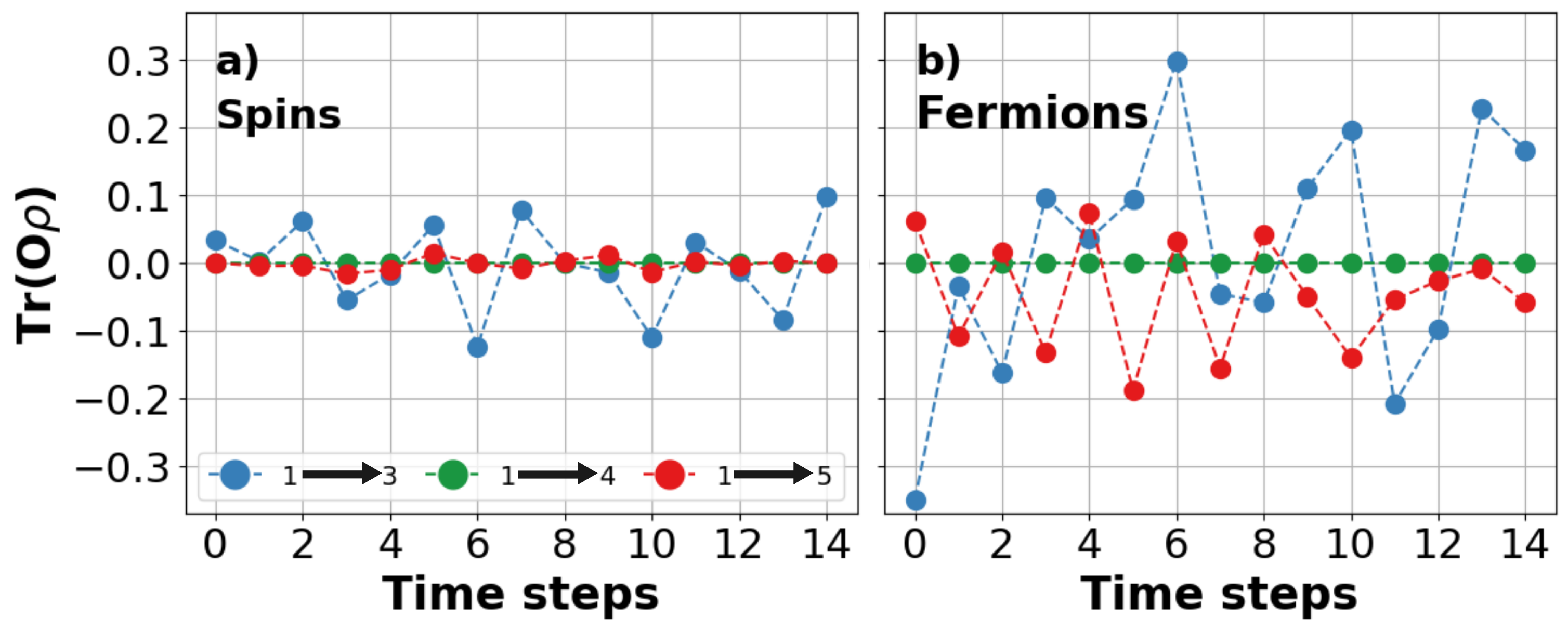}
    \caption{Time evolution of the expected values for non-diagonal observables with a one-dimensional chain of five particles, where the coupling constant is $J_{i, i+1}=1$. Injecting the same sequence of inputs into the first particle, (b) fermions are able to modify the state of the fifth particle with much more intensity than qubits (a). As a note, for symmetry reasons, these correlations vanish at odd-distant sites, e.g. $\langle a_1^{\dagger} a_4 + h.c.\rangle$(green line).} 
    \label{fig:1d-chain topology}
\end{figure}

So far, the dynamics of the reservoir was controlled by the Eq. \eqref{eq:quadratic_hamiltonian}, which allows the exchange of information between all particles. To simplify the dynamics, 
let us take a one-dimensional, nearest-neighbor chain of $N=5$ particles with homogeneous couplings 
\begin{equation}
     H=\sum_{i=1}^{N-1}a_{i}^{\dagger}a_{i+1} +a_{i+1}^{\dagger}a_{i}.
     \label{eq:1d-chain}
\end{equation}
With this 
simplified topology, at each iteration, we keep injecting information into particle $1$ and after a unitary evolution with $dt=10$, we measure the effect of 
input information spreading towards
different sites of the chain using non-diagonal observables, Fig. \ref{fig:1d-chain topology}. Without surprise, the impact of the input injection decays with the distance. Observables at distance $2$ display stronger correlations, $\langle a_1^{\dagger} a_3 + h.c. \rangle$ (blue) than observables at larger distance, e.g comparing with correlations at $4$ sites separation,  $\langle a_1^{\dagger} a_5 + h.c.\rangle$ (red).
The relevant feature of Fig. \ref{fig:1d-chain topology} is the disparity between the expected value of qubits and fermions. Since at a distance of $4$ (red) the expected value of fermions is greater than qubits, we have a strong indication about the ability of fermions to delocalize way information more efficiently than qubits.

\begin{figure}[h]
    \centering
    \includegraphics[width=0.48\textwidth]{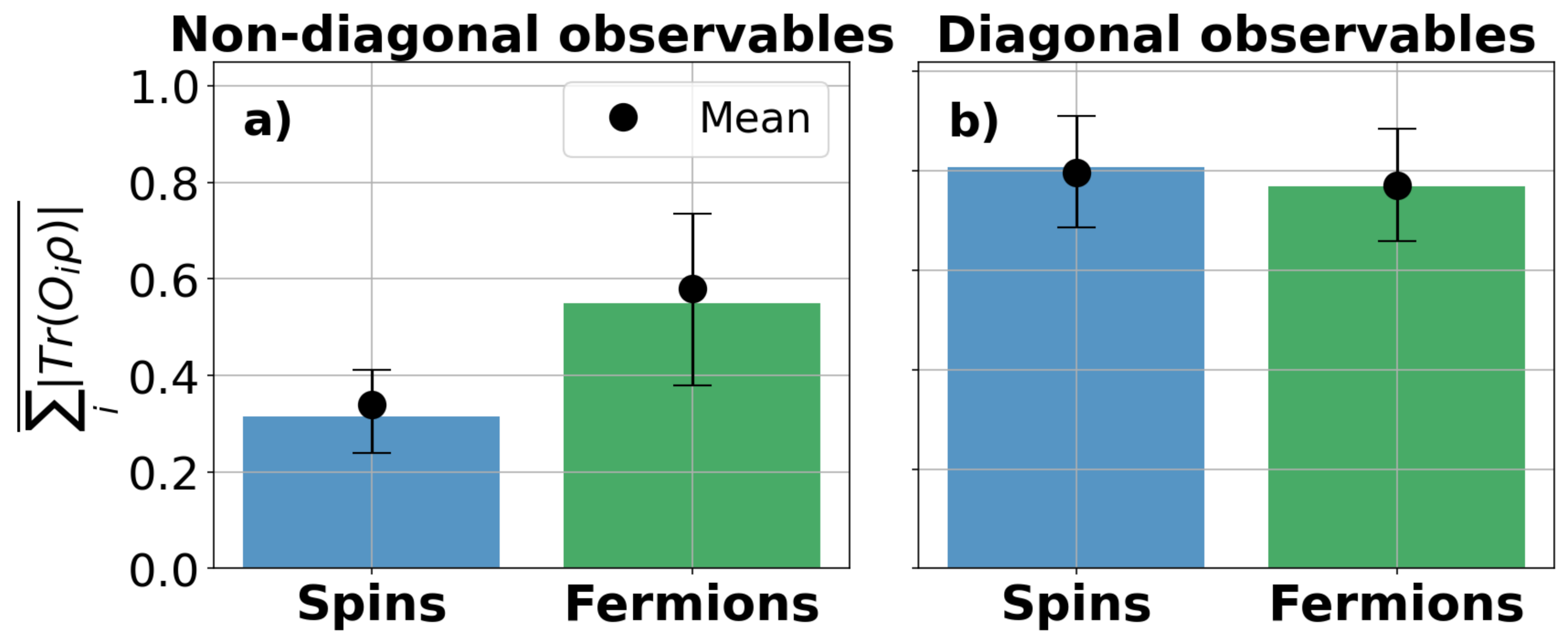}
    \caption{Sum over the absolute value of all (a) non-diagonal and (b) diagonal observables at an instant of time. Quantitatively, the results will change for different instants of time, but the qualitative behavior remains at each instant of time. The error bars correspond to the first and third quantiles over 1000 realizations of $J_{ij}$. As in Fig. \ref{fig:1d-chain topology}, fermions are able to store more information on nonlocal terms than qubits (a). In contrast, no relevant differences are found looking at the diagonal observables.}
    \label{fig:all-to-all topology}
\end{figure}

This effect is harder to detect if the topology of the reservoir is all-to-all Eq. \eqref{eq:quadratic_hamiltonian},
so that the expected value will not decay with the distance,  because all particles are directly connected with particle $1$ where injection occurs. 
Still, for random network reservoirs, we can sum over all ``non-diagonal'', that is, two-site observables.  
Since our objective is to see the amount of information stored in all these observables, we sum their absolute values.
Fig. \ref{fig:all-to-all topology} provides then insight to interpret why fermions have a higher memory capacity than qubits. Although in the diagonal (single-site) terms, qubits reach slightly higher values   than fermions, the relevant difference is in the non-diagonal terms, where fermions surpass qubits again. The anticommutativity nature of fermions plays a crucial role in how the system processes information and can lead to better performance, at least in temporal tasks, corroborating our interpretation. 

While the previous analysis allows us to make the point about the difference between qubits and fermions, the extension to the case of bosons would be less transparent, as the number of possible observables that one can build is much larger. For instance, one could look at operators like $ (a_i^{\dagger n} a_j^{m}+h.c.)$. Then, the absolute values of the hopping operators used in Fig. \ref{fig:1d-chain topology} would not be representative of the complexity of the boson dynamics, even in the few-excitation regime.

\section{Conclusions}\label{sec:conclusions}

In this work, we have studied the performance of QRC using the lens of quantum statistics by comparing the processing capabilities for the cases of fermions, spins, and bosons. 
While good information processing capacity can be achieved with any system,  and even for the simple hopping Hamiltonian considered here, we have identified fundamental differences that arise from the dynamics of each kind of particle. The main factors determining the QRC  performance are the Hilbert space size, the number of available observables limiting the size of the output layer, and the information-spreading proclivity of the particles involved. Bosons potentially offer a key advantage due to the infinite size of the corresponding Hilbert space. Still, it does not translate into an appealing feature in numerical QRC, as populating high-energy levels would require higher computational power. 

In experimental realizations, where this is not an issue, an interesting question is a need for a more complex dynamics to avoid linear dependence between different observables.  
For example, it is known that considering the linear dynamics in bosonic systems limited to Gaussian states allows the achievement of a polynomial advantage, but not an exponential one
\cite{nokkala2020gaussian}.
Also, recent results with a single qudit display a saturation in the performance as the number of levels increases, which is a spy of the lack of linear independence between the available output observables \cite{Kalfus2022}.

A less expected result is that, for a given size of the Hilbert space (limiting bosons to few excitations), fermions are better information vectors due to the anticommutation rules, at least for linear tasks. This can be seen clearly by comparing them to qubits, with which the ``sole" difference is exactly given by nonlocal commutation rules. 
We have also shown that a tailored input injection strategy can be very efficient not only  to change  the linear/nonlinear response of the reservoir \cite{mujal2021analytical,Govia_2022} but also to enhance the whole performance of the reservoir.
Our results also  show  that, when bosonic levels higher than 1 are not populated, the QRC performance approaches the one obtained with spins.   

An important point, which has not been touched on in this paper and that is related to the statistics transition driven by a Hamiltonian parameter,  concerns the presence of non-quadratic terms explicitly in the Hamiltonian \cite{2019Ghosh,PhysRevLett.123.260404}.   How does such a nonlinearity affect the nonlinear input-output QRC map? Is there any difference due to the quantum commutation rules? The class of models admitting such nonlinear terms could also be suited to study  the possibility of exploring different dynamical regimes, like the ones found in the systems studied in Refs. \cite{buch21,buch22}, and working near the edge of instability, drawing analogies with the results of Ref. \cite{2021Martinez}.

\section*{Acknowledgements}
We acknowledge useful discussions with M. C. Soriano and F. Plastina.
Funding acknowledged from the Spanish State Research Agency, through  the  QUARESC project (PID2019-109094GB-C21/AEI/ 10.13039/501100011033) and the Severo Ochoa and Mar\'ia de Maeztu Program for Centers and Units of Excellence in R\&D (MDM-2017-0711), from CAIB through the QUAREC project (PRD2018/47), and from the CSIC Interdisciplinary Thematic Platform (PTI+) on Quantum Technologies in Spain (QTEP+).
GLG is funded by the Spanish  MEF/MIU and co-funded by the University of the Balearic Islands through the Beatriz Galindo program (BG20/00085). CC was supported by Direcció General de Política Universitària i Recerca from the government of the Balearic Islands through the postdoctoral program Margalida Comas. We also want to thank Tomoyuki Kubota for sharing his code on \url{https://github.com/kubota0130/ipc}.

\section*{Data Availability Statement}
The numerical results presented in this paper can be reproduced using the codes available at \url{https://github.com/gllodra12/Benchmarking_QRC}.

\appendix
\renewcommand{\thefigure}{A\arabic{figure}}
\setcounter{figure}{0}
\section{Dynamics response to the input injection}\label{appendix:dynamics}

In the main text, we reported the performance of the memory capacity with different observables, see Fig. \ref{fig:linear_observables}. The MC is found to vanish when the dynamics of the reservoir does not capture the necessary features to distinguish between different inputs. 

\begin{figure}[h]
    \centering
    \includegraphics[width=0.48\textwidth]{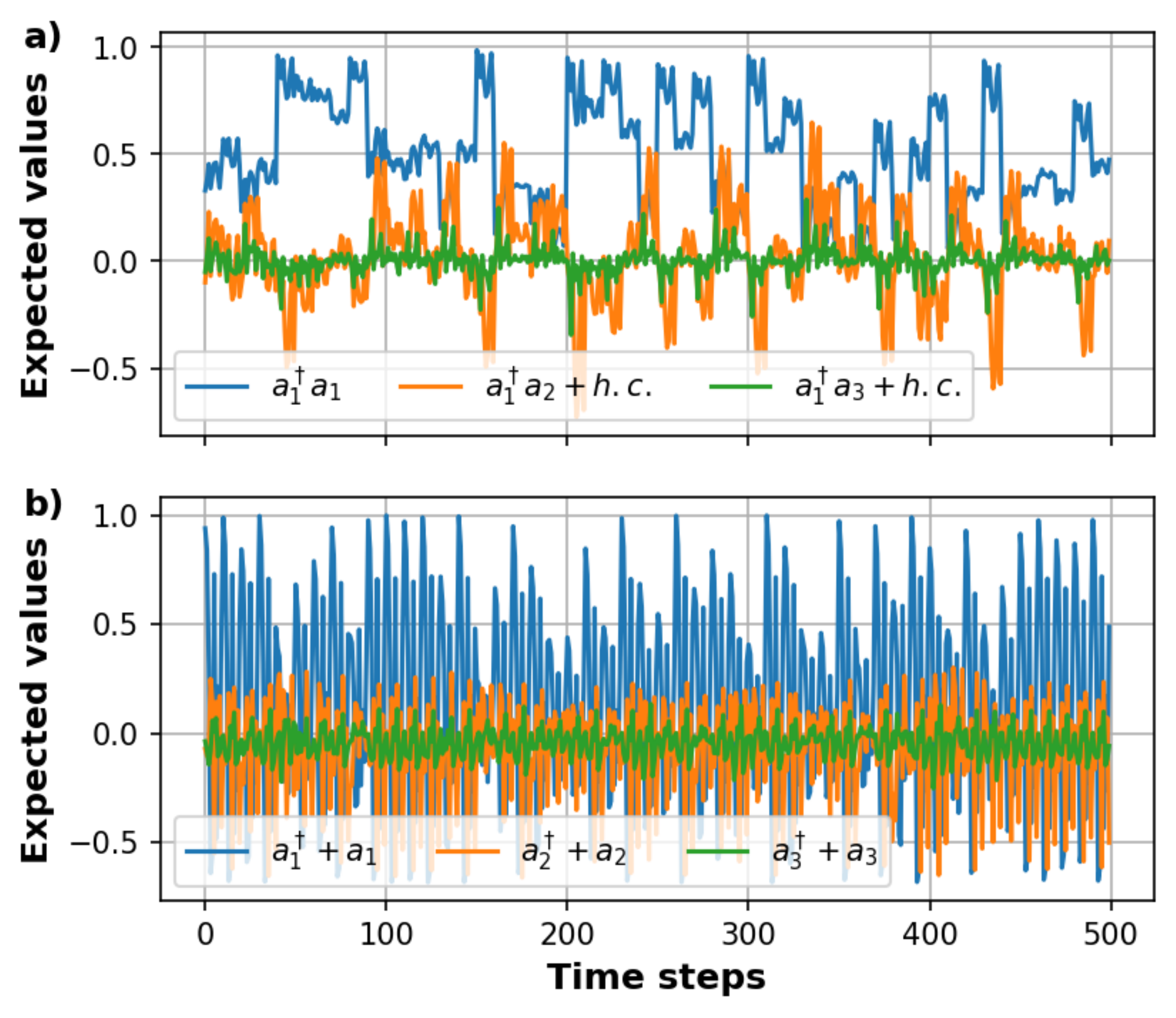}
    \caption{Dynamics of fermionic observables $\langle a^{\dagger}_i a_j\rangle+h.c.$ and $\langle a^{\dagger}_i + a_i\rangle$ for one random realization. Every 10 time steps, we inject a random input. (a) The dynamics of the observable $\langle a^{\dagger}_i a_j\rangle+h.c.$ is driven by an input injected into particle 1. (b) However, for $\langle a^{\dagger}_i + a_i\rangle$ the input, show little correlation with respect to the dynamics. For other random realizations, the qualitative aspect of the dynamics doesn't change.}
    \label{appendix_fig:fermion_dynamics_obs_ij_and_i+i}
\end{figure}

In Fig. \ref{appendix_fig:fermion_dynamics_obs_ij_and_i+i} we compare  the dynamics of different observables. In particular, we plot in Fig. \ref{appendix_fig:fermion_dynamics_obs_ij_and_i+i}a observables of the kind $\langle a_{i}^{\dagger}a_{j}\rangle+h.c.$, which we know from Fig. \ref{fig:linear_observables} that guarantee good memory,  and in Fig. \ref{appendix_fig:fermion_dynamics_obs_ij_and_i+i}b observables of the form $\langle a_{i}^{\dagger}+a_{i}\rangle$, which give almost zero memory capacity. Indeed, looking at these figures, it is clear that in the latter case, the dynamics is almost insensitive to new input injections (which take place every ten timesteps), while in the former case the response  is evident.

\renewcommand{\theequation}{B\arabic{equation}}
\setcounter{equation}{0}
\renewcommand{\thefigure}{B\arabic{figure}}
\setcounter{figure}{0}
\section{Information processing capacity}\label{appendix:IPC}
The information processing capacity (IPC) is a measure proposed in Ref. \cite{2012Dambre} and extended in \cite{kubota2021unifying} to quantify the computational capabilities of a dynamical system by evaluating all the possible linear and nonlinear contributions to the memory. For instance, in the main text, we have studied tasks of the form $\hat{y}_k=u_{k-\tau}^q$, which are a subset of all nonlinear tasks. Obviously, it is impractical to study all nonlinear tasks. For this reason, the IPC evaluates the normalized Legendre polynomials, which form a complete set of orthogonal functions. So, the IPC studies tasks of the form: 

\begin{equation}
    \hat{y}_k = \prod_{\tau}\mathcal{P}_{q}(s_{k-\tau}),
    \label{eq:target_ipc}
\end{equation}
where $\mathcal{P}_q(\cdot)$, $q \geq 1$ is the normalized Legendre polynomial of degree $q$ and $s_{k-\tau}$ is the sequence of inputs over the interval $[-1, 1]$ 
(notice the difference with the kind of input encoding introduced in Eq. \eqref{eq:input_state} of the main text). 
Then, in this case, we redefine the input injection state as
\begin{equation}
|\psi_{k}^{(e)}\rangle =\sqrt{\frac{1}{2}(1+s_k)}|0\rangle +\sqrt{\frac{1}{2}(1-s_k)}|e\rangle. 
\label{eq:input_ipc}
\end{equation}
To quantify how well a system can reproduce the target function defined in Eq. \eqref{eq:target_ipc}, the capacity coefficient is
\begin{equation}
    C(\textbf{X}, y)=1-\frac{min_w \text{MSE}(y, \hat{y})}{\langle\hat{y}^2\rangle},
    \label{eq:capacity_ipc}
\end{equation}
where $\textbf{X}$ is the state that describes the reservoir, $y$ and $\hat{y}$ are the predictions and target outputs. Finally, $w$ is the weight vector of the output layer. The mean square error (MSE) is the cost function, $\text{MSE}(y, \hat{y})=\frac{1}{L}\sum_{k=1}^L(y_k-\hat{y}_k)^2$, that evaluates if the prediction is close to the target output. As a normalization factor, the bracket $\langle \cdot \rangle$ denotes the average over the target output. 

One of the main results of \cite{2012Dambre} is that the total capacity, $C_T=\sum_q C(\textbf{X}, y(q))$, which is the sum over all degrees (linear $q=1$ and nonlinear $q\geq1$), is upper-bounded by the number of linearly independent variables of the output layer. This theoretical bound requires that the system satisfies the fading memory, which is strongly connected to the convergence property. 

\begin{figure}[h]
    \centering
    \includegraphics[width=0.48\textwidth]{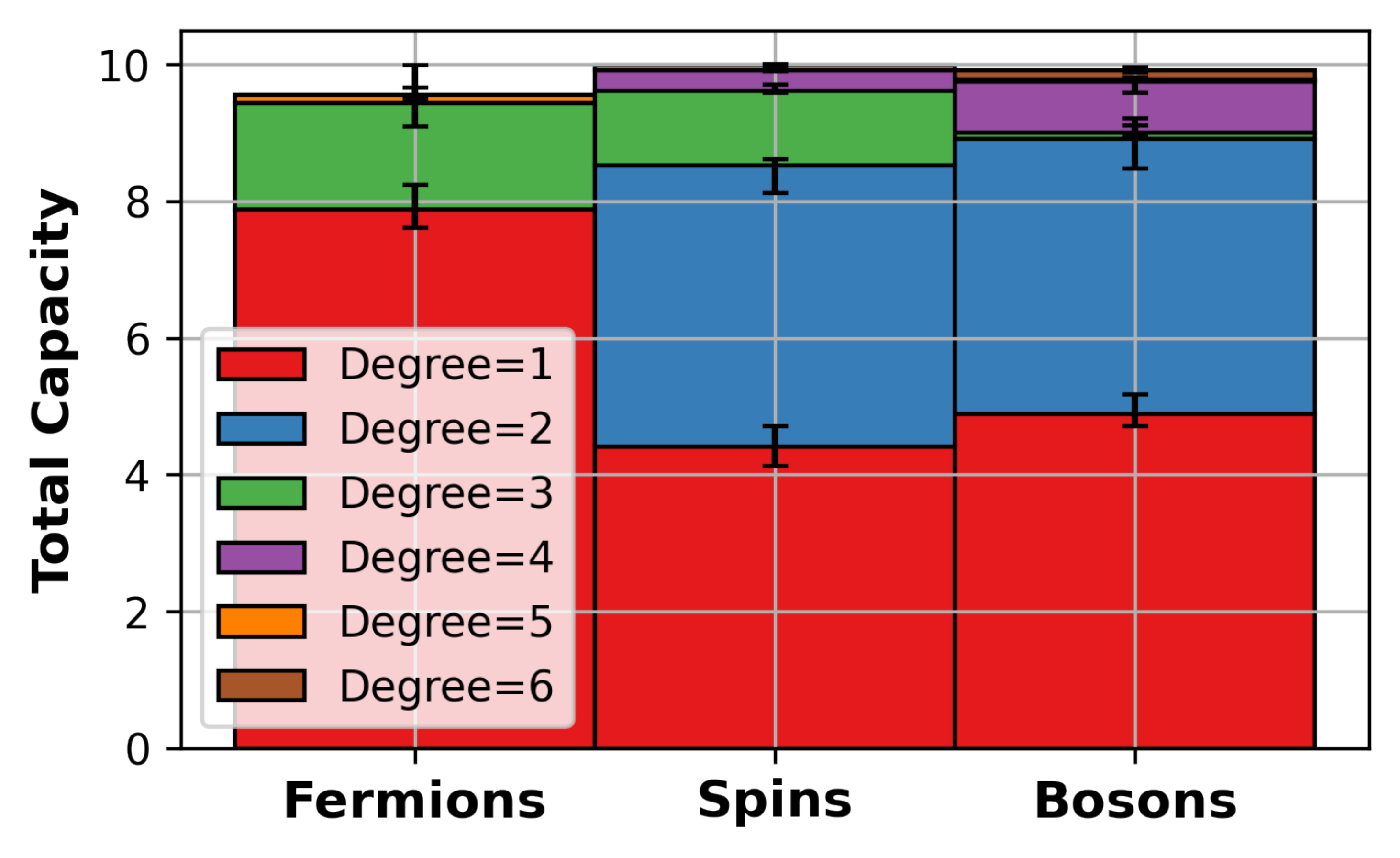}
    \caption{Information processing capacity for different particles. Fermions are not able to process Legendre polynomials of even degree. In contrast, spins and bosons are able to handle all Legendre polynomials, even though  bosons have  poor performance with odd degrees.
    In all particles, the training set has been increased up to $L=10^6$ to reduce the level of noise, and the maximum degree of nonlinearity $q_{max}=6$ is sufficient to guarantee the saturation of the total capacity in our case. The total capacity is bounded at ten because we have four particles and the set of observables are of the form $a^{\dagger}_i a_j$ ($M=10$).
    The height of each column corresponds to the median and the error bars to the quantiles Q25 and Q75 over 10 realizations.}
    \label{appendix_fig:total_capacity}
\end{figure}

In Fig. \ref{appendix_fig:total_capacity} we show the total capacity for fermions, spins, and bosons with an output layer of ten observables. We can see that the overall IPC is similar in all cases, even though spins saturate the theoretical bound while bosons and especially fermions remain a bit below such a bound. This is caused by the fact that some of the realizations do not satisfy the fading memory condition. In particular for fermions, even though most realizations saturate the upper bound, there are a few of them that do not saturate. Unsurprisingly, the realizations that do not reach the upper bound are the ones that have a low convergence.  This small difference in the total IPC is not in contradiction with the results of the main text, where it is shown the behaviors of different particles in specific, more realistic tasks.

\begin{figure}[h]
    \centering
    \includegraphics[width=0.48\textwidth]{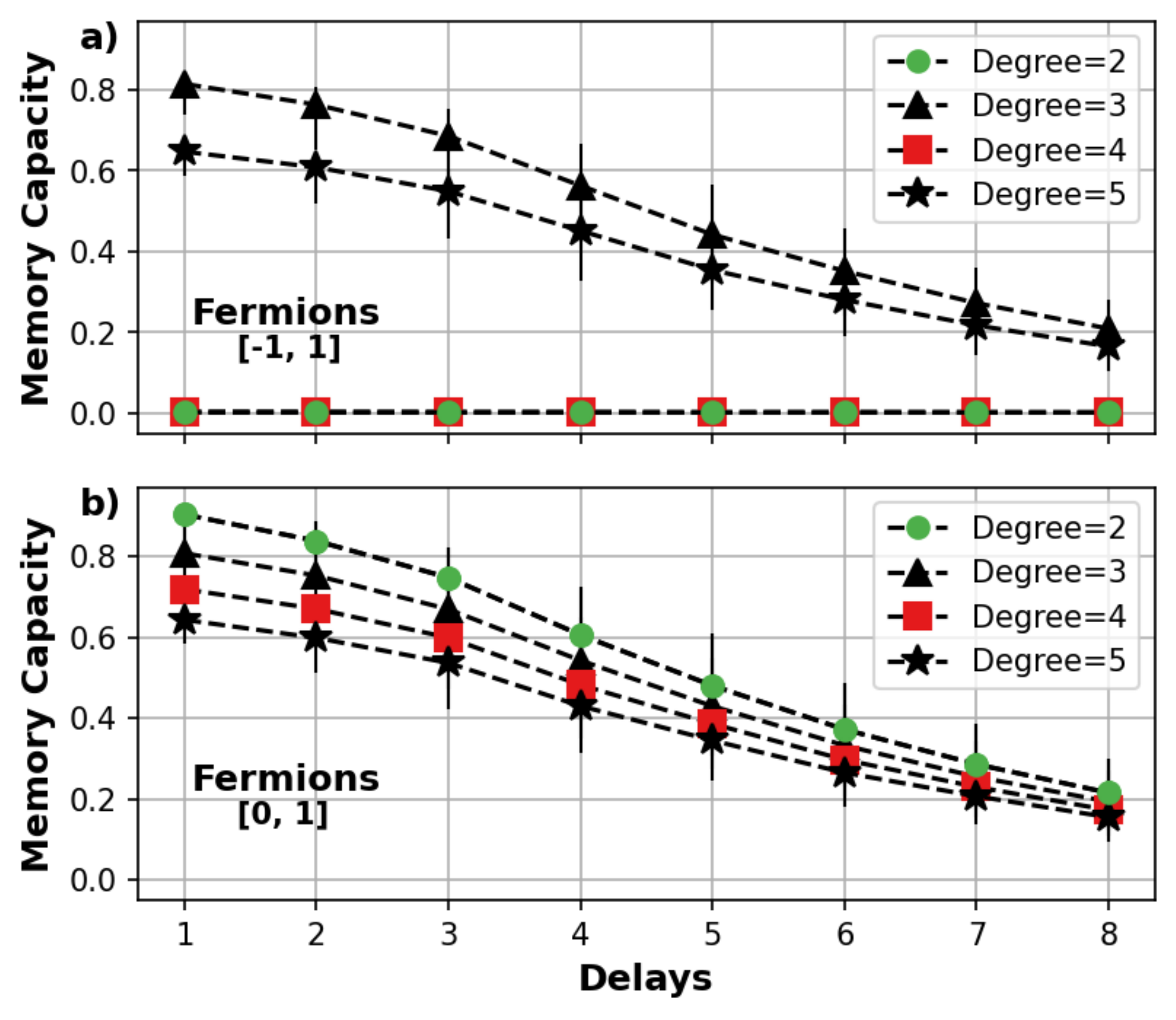}
    \caption{Comparing different inputs intervals, $s_k$ and $u_k$, respectively. (a) As the IPC in Fig. \ref{appendix_fig:total_capacity} points out, fermions have low memory capacity for even degrees. However, if the input encoding is $u_k$ (b) fermions are able to process tasks with even or odd degrees.}
    \label{appendix_fig:mc_input_-1,+1_and_0,1}
\end{figure}

An interesting result of Fig. \ref{appendix_fig:total_capacity} concerns the fact that fermions look unable  to process functions with an even degree of nonlinearity. This apparent  contradiction with respect to Fig. \ref{fig:non_linear}d of the main text is originated from the different form of input injection used to calculate the IPC in Eq. \eqref{eq:input_ipc} with respect to Eq. \eqref{eq:input_state} (a deep discussion about how nonlinear functions of the input propagate into the output can be found in \cite{mujal2021analytical}). 
For the sake of completeness, in Fig.   \ref{appendix_fig:mc_input_-1,+1_and_0,1}  we reproduce the MC of fermions for different degrees of nonlinearity considering both input functions. While for odd degrees of nonlinearities there is no appreciable difference,  in the case of even degrees, we observe the behavior described above using the IPC. 

To conclude, both MC and IPC are valuable metrics to analyze how a dynamical system process information, although the IPC is time expensive to compute in comparison to the MC. Both metrics complement each other quite well since they offer distinct points of view on the processing capacity of a reservoir computer.

\newpage
\bibliographystyle{unsrt}
\bibliography{QRCbib}

\end{document}